\documentclass[journal]{IEEEtran}
\ifCLASSINFOpdf
\else
\fi
\usepackage{nicefrac}       
\usepackage{amsfonts,amsthm,amsopn,mathrsfs}
\usepackage{algpseudocode}
\usepackage[Algorithmus]{algorithm}
\usepackage{enumitem,cite,url,bm,amsmath,amssymb,graphicx,subfigure,enumitem,float,pifont,xfrac,mathtools}
\usepackage{multirow}
\usepackage{newtxmath}

\usepackage{hyperref}
\hypersetup{colorlinks=black, pdfstartview=FitV, linkcolor=blue, citecolor=black, plainpages=false, pdfpagelabels=true, urlcolor=blue}
\usepackage[all]{hypcap}

\usepackage{color}
\usepackage{dcolumn}

\definecolor{00BCB4}{RGB}{0,188,180}
\definecolor{C4E86B}{RGB}{196,232,107}
\definecolor{49DEB2}{RGB}{73,222,178}
\definecolor{FF4747}{RGB}{255,71,71}
\definecolor{FF3561}{RGB}{255,53,97}
\definecolor{CD2C6C}{RGB}{205,44,108}
\definecolor{10B48E}{RGB}{16,180,142}
\definecolor{8FDBD8}{RGB}{143,219,216}

\usepackage{etoolbox}
\makeatletter
\patchcmd{\maketitle}
 {\def\@makefnmark}
 {\def\@makefnmark{}\def\useless@macro}
 {}{}
\makeatother

\usepackage{tikz}
\usetikzlibrary{bayesnet}
\usetikzlibrary{arrows,shapes,snakes,automata,backgrounds,petri}

\newcommand{\vw}{{\bm w}}
\newcommand{\vx}{{\bm x}}
\newcommand{\vy}{{\bm y}}
\newcommand{\vz}{{\bm z}}

\newcommand{\vq}{{\bm q}}

\newcommand{\mA}{{\bm A}}

\newcommand{\mI}{{\bm I}}

\usepackage{chngcntr}
\usepackage{apptools}
\AtAppendix{\counterwithin{lemma}{section}}

\newenvironment{sparse_signal_model}[1][\ding{113} Sparse signal model:\\]{\begin{trivlist}\item[\hskip \labelsep {\bfseries #1}]}{\end{trivlist}}
\newenvironment{1_bit_quantization_noise_model}[1][\ding{113} 1-bit quantization noise model:\\]{\begin{trivlist}\item[\hskip \labelsep {\bfseries #1}]}{\end{trivlist}}

\algnewcommand{\IIf}[1]{\State\algorithmicif\ #1\ \algorithmicthen}
\algnewcommand{\EndIIf}{\unskip\ \algorithmicend\ \algorithmicif}

\allowdisplaybreaks

\begin{document}
%
\title{1-Bit Compressive Sensing via Approximate Message Passing with Built-in Parameter Estimation}
%
%
%

\author{Shuai Huang, and Trac D. Tran, ~\IEEEmembership{Fellow,~IEEE}
\thanks{This work is supported by the National Science Foundation under grants NSF-CCF-1117545, NSF-CCF-1422995 and NSF-ECCS-1443936.}
\thanks{The authors are with the Department of Electrical and Computer Engineering, Johns Hopkins University, Baltimore, MD, 21218 USA (email: shuai.huang@emory.edu; trac@jhu.edu).}}

\maketitle

\begin{abstract}
1-bit compressive sensing aims to recover sparse signals from quantized 1-bit measurements. Designing efficient approaches that could handle noisy 1-bit measurements is important in a variety of applications. In this paper we use the approximate message passing (AMP) to achieve this goal due to its high computational efficiency and state-of-the-art performance. In AMP the signal of interest is assumed to follow some prior distribution, and its posterior distribution can be computed and used to recover the signal. In practice, the parameters of the prior distributions are often unknown and need to be estimated. Previous works tried to find the parameters that maximize either the measurement likelihood via expectation maximization, which becomes increasingly difficult to solve in cases of complicated probability models. Here we propose to treat the parameters as unknown variables and compute their posteriors via AMP as well, so that the parameters and the signal can be recovered jointly. Compared to previous methods, the proposed approach leads to a simple and elegant parameter estimation scheme, allowing us to directly work with 1-bit quantization noise model. Experimental results show that the proposed approach generally perform much better than the other state-of-the-art methods in the zero-noise and moderate-noise regimes, and outperforms them in most of the cases in the high-noise regime.
\end{abstract}


%
\IEEEpeerreviewmaketitle

\section{Introduction}
Compressive sensing (CS) has enabled us to recover a signal with prior information at lower sampling rates \cite{Decode05,RUP06,CS06,SRRP06}. Sparse signal recovery is the key topic in compressive sensing that lays the foundation for applications such as dictionary learning \cite{DL06,Denoising06}, sparse representation-based classification \cite{SRC09}, channel estimation \cite{Berger:UW:2010,Berger:CE:2010,Bajwa:2010}, etc. Here we would like to recover a sparse signal $\vx\in\mathbb{R}^N$ given the measurement matrix $\mA\in\mathbb{R}^{M\times N}$ and quantized measurements $\vy$
\begin{align}
\label{eq:quantized_measurements}
\vy=\mathscr{Q}(\mA\vx+\vw)\,,
\end{align}
where $\mathscr{Q}(\cdot)$ is the quantization operator, and $\vw$ is the noise. The problem itself is generally ill-posed, and we rely on the prior information that the signal is sparse to recover it. In the extreme case where $\vy\in\{-1,+1\}^M$, we have the \emph{``1-bit compressive sensing''} problem originally proposed in \cite{Boufounos:1bitCS:2008}. It arises from applications such as channel estimation in the massive multiple-input-multiple-output (MIMO) system, where the channel matrix is approximately sparse in the angle domain, and linear measurements are acquired using analog-to-digital converters (ADCs) \cite{Larsson:MIMO:2014,Wen:Channel:2015}. 

In this case, power consumption of the ADCs grows exponentially with the number of quantization bits, along with the drastically increased cost and difficulty in hardware design \cite{Walden:ADC:1999}. Currently these issues make it either too expensive or impractical to deploy high-resolution ADCs in base stations and portable devices \cite{Murmann:1997-2020}. As a result, there has been a growing interest in low-resolution ADCs that output $1\sim4$ bits. Considerable efforts have been made to perform channel estimation from low-bit measurements in recent years \cite{Wen:LowADC:2016,Jacobsson:LowADC:2017,Mo:LowADC:2018}. In particular, 1-bit ADCs are much preferred in wideband millimeter wave communication systems that require high sampling frequency \cite{Rappaport:MmWave:2015,Mo:1bitADC:2015,Jacobsson:1bitADC:2015,Choi:1bitADC:2016,L1:1bitADC:2016}.

Depending on how the sparse prior is enforced, various approaches have been proposed to solve the 1-bit CS problem. Both the binary iterative hard thresholding (BIHT) algorithm \cite{Jacques:1bitCS:2013} and the convex programming approach \cite{Plan:1bitCS:2013} impose a constraint on the sparsity of the signal, i.e. $\|\vx\|_0\leq K$, where $K$ is the number of nonzero entries in $\vx$. A linear programming formulation that minimizes the $l_1$-norm $\|\vx\|_1$ subject to the convex constraints from noiseless measurements can be also derived \cite{Plan:1bitCS:2013:2}. 

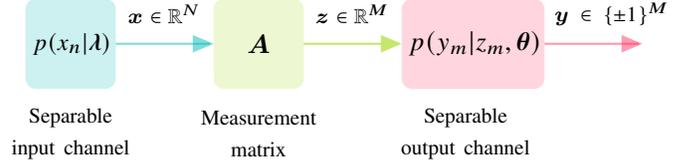
\begin{figure}[tbp]
\begin{tabular}{c}
\hspace*{-2.5em}
\definecolor{00BCB4}{RGB}{0,188,180}
\definecolor{C4E86B}{RGB}{196,232,107}
\definecolor{FF3561}{RGB}{255,53,97}

\begin{tikzpicture}

  \fill[00BCB4, opacity=0.2, rounded corners] (0,0) rectangle (1.2,1.2);
  \node[text width=1cm, align=center] at (0.6,0.6) {\small$p(x_n|\boldsymbol\lambda)$};
  \node[text width=2cm, align=center] at (0.6, -0.6) {\footnotesize Separable input channel};
  
  \draw[00BCB4, opacity=0.5, thick, ->] (1.2,0.6) -- (2.5,0.6) ;
  \node[text width=1cm, align=center] at (1.85, 1) {\footnotesize $\vx\in\mathbb{R}^N$};
  
  \fill[C4E86B, opacity = 0.3, rounded corners] (2.5,0) rectangle (3.7,1.2);
  \node[text width=1cm, align=center] at (3.1,0.6) {$\mA$};
  \node[text width=2cm, align=center] at (3.1,-0.6) {\footnotesize Measurement matrix};
  
  \draw[C4E86B, opacity = 1, thick, ->] (3.7,0.6) -- (5,0.6);
  \node[text width=1cm, align=center] at (4.35, 1) {\footnotesize $\vz\in\mathbb{R}^M$};
  
  \fill[FF3561, opacity=0.2, rounded corners] (5,0) rectangle (6.9,1.2);
  \node[text width=1cm, align=center] at (5.6, 0.6) {$p(y_m|z_m,\boldsymbol\theta)$};
  \node[text width=2cm, align=center] at (5.85, -0.6) {\footnotesize Separable output channel};
  
  \draw[FF3561, opacity=0.5, thick, ->] (6.9,0.6) -- (8.2, 0.6);
  \node[text width=2.5cm, align=center] at (7.8, 1) {\footnotesize $\vy\in\{\pm1\}^M$};

\end{tikzpicture}
\end{tabular}
\caption{A probabilistic view of the sparse signal recovery \cite{Rangan:GAMP:2011}: the signal $\vx$ follows a prior distribution $p(x_n|\boldsymbol\lambda)$, the noiseless measurements $\vz=\mA\vx$ are further corrupted by noise, producing the noisy measurements $\vy$ that follow the distribution $p(y_m|z_m,\boldsymbol\theta)$. The distribution parameters $\boldsymbol\lambda,\boldsymbol\theta$ are unknown and need to be estimated.}
\label{fig:bayesian_model}
\end{figure}

On the other hand, the sparse prior can be interpreted from a probabilistic perspective. Under the Bayesian setting shown in Fig. \ref{fig:bayesian_model}, belief propagation \cite{Pearl:1988,Kschischang:2001,Koller:2009}, also known as the sum-product message passing, can be used to perform probabilistic inference on the factor graph of the sparse signal recovery task \cite{Donoho:AMP:2009,Baron:2010,Rangan:GAMP:2011,Krzakala:2012:1}. Exact message passing is generally difficult to compute, and approximated message passing (AMP) is often used instead. The distributions are approximated by a family of simple distributions such as the Gaussians in AMP, where some chosen divergence measure such as the Kullback-Leibler divergence between the true distribution and the approximated distribution is minimized \cite{Minka:2001,Minka:Div:2005,Wainwright:Graph:2008}. AMP is computationally efficient and achieves state-of-the-art performances in channel estimation from low-bit measurements in the massive MIMO systems \cite{Wen:Channel:2015,Wen:LowADC:2016,Mo:LowADC:2018,Bellili:Lap:2019,Myers:1bit:2019}. 

The parameters of the prior distributions in AMP are unknown in practice and need to be estimated in order to recover the signals. Inspired by the long history of treating the distribution parameters as random variables in mathematical statistics \cite{Bickel:2015}, we proposed an extension to the AMP framework in \cite{PE_GAMP17} where the posteriors of the signal and parameters can be computed and used to recover them jointly. In this paper we present a more computationally efficient approach to perform parameter estimation, and show that the quantization noise model in 1-bit compressive sensing can be handled with ease. Experimental results show that the proposed approach generally perform much better than the other state-of-the-art methods, and is able to match the performance of the oracle AMP where true distribution parameters are known.

\subsection{Prior Art}
AMP was first used to solve large-scale compressive sensing problems in \cite{Donoho:AMP:2009}. Its Bayesian formulations in the form of belief propagation were later introduced in \cite{Rangan:GAMP:2011,Krzakala:2012:1}. In this paper we shall adopt the formulation in \cite{Rangan:GAMP:2011} termed ``generalized approximate message passing'' (GAMP). In order to estimate the distribution parameters, the measurement likelihood is maximized in \cite{Vila:BG:2011,Vila:EMGM:2013}, and Bethe free entropy \cite{Marc:IPC:2009} is maximized in \cite{Krzakala:2012:1,Krzakala:2012:2}. However, the computations involved in both approaches become increasingly difficult in the case of complicated probability models such as the quantization noise model. As a result, AWGN model was adopted in \cite{risi2014massive, Wang:Multiuser:2015,Wen:Channel:2015,Bellili:Lap:2019} to approximate the quantization noise model, which leads to sub-optimal performance. The approach in \cite{Mo:LowADC:2018} adopted the true quantization noise model but assumed the noise distribution parameter was already known, i.e. it needed to be manually tuned. By treating the parameters as unknown random variables and maximizing their posteriors, we show that our proposed approach has a wider applicability and could directly work with the complicated quantization noise model.

Using AMP to solve the 1-bit compressive sensing problem has been studied with different quantization noise models in \cite{Kamilov:1bit:2012,Yang:AMPquan:2013,Musa:GAMP1bit:2016,Movahed:1bit:2016,Kafle:1bit:2019}. The noise distribution parameters are either prespecified or need to be tuned manually. Let $\vw$ denote the white Gaussian noise $\vw\sim\mathcal{N}(\boldsymbol0,\gamma_w\mI)$. The noise $\vw$ is added before the quantization in \cite{Jacques:1bitCS:2013,Yang:AMPquan:2013,Movahed:1bit:2016} where $\vy=\textrm{sign}(\mA\vx+\vw)$, whereas $\vw$ is added after the quantization in \cite{Musa:GAMP1bit:2016} where $\vy=\textrm{sign}(\mA\vx)+\vw$. In this paper we work with quantized measurements from ADC whose input-referred noise is added before the quantization. 

When the measurement matrix $\mA$ is i.i.d. zero-mean Gaussian, the convergence behavior of AMP in the large system limit can be characterized by the state evolution that predicts how the variables evolve through the iterations \cite{Donoho:AMP:2009,Bayati:SE:2011}. The GAMP formulation adopted in this paper also agrees with the state evolution \cite{Rangan:GAMP:2011}, and consistent parameter estimation can be guaranteed \cite{Kamilov:PE:2014}. A new belief propagation formulation termed ``vector approximate message passing'' (VAMP) was proposed in \cite{Rangan:VAMP:2017}, and its state evolution applies to a broader class of random matrices $\mA$ that are right-orthogonally invariant. Although it is still an open problem as to how the state evolution analysis can be derived for more general measurement matrices, AMP has been used with empirical success in real applications like channel estimation \cite{Wen:Channel:2015,Wen:LowADC:2016,Mo:LowADC:2018} and phase retrieval \cite{Schniter:PR:2015,Metzler:PR:2015}. In practice, operations like damping and mean removal are quite effective in preventing divergence of the algorithm \cite{Rangan:DampingCvg:2014,Vila:DampingMR:2015}.

\subsection{Main Contribution and Paper Outline}
Inspired by the practice of treating distribution parameters as unknown variables in mathematical statistics \cite{Bickel:2015}, we perform parameter estimation in a much simpler manner by maximizing the posteriors of the parameters in AMP. This allows us to consider complicated probability models under the AMP framework. Building upon our earlier work in \cite{PE_GAMP17}, we propose a more computationally efficient approach that combines expectation maximization (EM) \cite{Dempster:EM:1977} and the second-order method in this paper, and use it to solve the 1-bit compressive sensing problem. Compared to previous AMP approaches that either prespecify/tune the parameters or use an approximated noise model, our approach directly works with the true 1-bit quantization noise model, which leads to much improved performances. 

This paper proceeds as follows. In Section \ref{sec:amp_pe_intro} we introduce the extended AMP framework that treats distribution parameters as unknown variables, and derive the messages passed among variable nodes. In Section \ref{sec:1bitCS} we present the sparse signal model and the 1-bit quantization noise model, and show how the distribution parameters can be estimated efficiently by maximizing their posteriors. We compare the proposed approach with the other state-of-the-art methods in Section \ref{sec:exp}. We finally conclude this paper with a discussion in Section \ref{sec:con}.

\section{AMP with Built-in Parameter Estimation}
\label{sec:amp_pe_intro}
\begin{figure}[tbp]
\begin{center}
\begin{tabular}{c}
\hspace*{-2em}
%
%
%
%
\definecolor{00BCB4}{RGB}{0,188,180}
\definecolor{C4E86B}{RGB}{196,232,107}
\definecolor{FF3561}{RGB}{255,53,97}

\begin{tikzpicture}

  
  \fill[00BCB4, opacity=0.2, rounded corners] (-0.7,-4.5) rectangle (2.5,1.9);
  \fill[C4E86B, opacity=0.3, rounded corners] (1.2,-3.4) rectangle (5.8,3);
  \fill[FF3561, opacity=0.2, rounded corners] (4.6,-4.5) rectangle (7.9,1.9);

  \fill[00BCB4, opacity=0.4, rounded corners] (-0.7,-4.5) rectangle (2.5,-3.5);
  \node[text width=3cm, align=center] at (0.9,-4) {Signal prior};
  \fill[C4E86B, opacity=0.7, rounded corners] (1.2,2) rectangle (5.8,3);
  \node[text width=4.5cm, align=center] at (3.5,2.5) {Measurement system};
  \fill[FF3561, opacity=0.4, rounded corners] (4.6,-4.5) rectangle (7.9,-3.5);
  \node[text width=3cm, align=center] at (6.25,-4) {Noise prior};

  \node[latent,fill=00BCB4!20] (lambda_1) {$\lambda_1$};
  \node[latent,fill=00BCB4!20, below = 1 of lambda_1] (lambda_L) {$\lambda_L$};
  
  \path (lambda_1) -- node[auto=false]{\vdots} (lambda_L);

  \node[latent,fill=C4E86B!30, above=0.533 of lambda_1, xshift=3.5cm] (x_1) {$x_1$};
  \node[latent,fill=C4E86B!30, below=0.666 of x_1] (x_2) {$x_2$};
  \node[latent,fill=C4E86B!30, below=2 of x_2] (x_N) {$x_N$};
  
  \path (x_2) -- node[auto=false]{\vdots} (x_N);
  
  \factor[left=0 of x_1, xshift=-1.5cm] {Omega_1-f} {above:$\Omega_1$} {x_1, lambda_1, lambda_L}{};
  \factor[left=0 of x_2, xshift=-1.5cm] {Omega_2-f} {above:$\Omega_2$} {x_2, lambda_1, lambda_L}{};
  \factor[left=0 of x_N, xshift=-1.5cm] {Omega_N-f} {below:$\Omega_N$} {x_N, lambda_1, lambda_L}{};
  \path(Omega_2-f) -- node[auto=false]{\vdots} (Omega_N-f);
  
  \node[latent,fill=FF3561!20, right=0 of lambda_1, xshift=6.5cm] (theta_1) {$\theta_1$};
  \node[latent,fill=FF3561!20, right=0 of lambda_L, xshift=6.5cm] (theta_K) {$\theta_K$};
  \path(theta_1) -- node[auto=false]{\vdots} (theta_K);
  

  \factor[right=0 of x_1, xshift=1.5cm] {Phi_1-f} {above:$\Phi_1$} {theta_1, theta_K, x_1, x_2, x_N} {};
  \factor[right=0 of x_2, xshift=1.5cm] {Phi_2-f} {above:$\Phi_2$} {theta_1, theta_K, x_1, x_2, x_N} {};
  \factor[right=0 of x_N, xshift=1.5cm] {Phi_M-f} {below:$\Phi_M$} {theta_1, theta_K, x_1, x_2, x_N} {};
  
  \path(Phi_2-f) -- node[auto=false]{\vdots} (Phi_M-f);
\end{tikzpicture}

\end{tabular}
\end{center}
\caption{The factor graph of the sparse signal recovery task: ``$\bigcirc$'' represents the variable node, and ``$\blacksquare$'' represents the factor node.}
\label{fig:factor_graph_pegamp}
\end{figure}
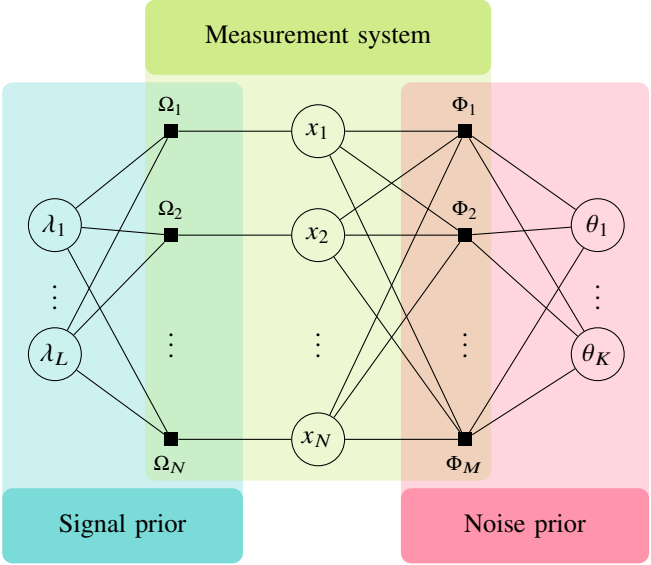

In this section we introduce the extended AMP framework from our earlier work \cite{PE_GAMP17} where the distribution parameters are treated as unknown variables. As shown in Fig. \ref{fig:factor_graph_pegamp}, the factor graph can be divided into three parts: the signal prior block that contains the signal distribution parameters $\boldsymbol\lambda=\{\lambda_1,\cdots,\lambda_L\}$, the measurement system block that contains the signal of interest $\vx=[x_1\ x_n \cdots\ x_N]^T$, and the noise prior block that contains the noise distribution parameters $\boldsymbol\theta=\{\theta_1,\cdots,\theta_K\}$. Inference tasks that compute the posteriors $p(\boldsymbol\lambda|\vy),\ p(\vx|\vy),\ p(\boldsymbol\theta|\vy)$ rely on the ``messages'' passed among different nodes. Taking the messages between the factor node $\Phi_m$ and the variable node $x_n$ for example, we use the following notations for the messages:
\begin{itemize}
\item $\Delta_{\Phi_m\rightarrow x_n}$ denote the message from $\Phi_m$ to $x_n$,
\item $\Delta_{x_n\rightarrow \Phi_m}$ denote the message from $x_n$ to $\Phi_m$.
\end{itemize}
Both $\Delta_{\Phi_m\rightarrow x_n}$ and $\Delta_{x_n\rightarrow \Phi_m}$ can be viewed as functions of $x_n$, and they are expressed in the ``$\log$'' domain in this paper.

Starting with the \emph{measurement system} block, we can write the messages exchanged among the nodes in the $(t+1)$-th iteration as follows:
\begin{subequations}
\label{eq:sp_fv_signal}
\begin{align}
\label{eq:sp_fv_phi_x}
\begin{split}
&\Delta^{(t+1)}_{\Phi_m\rightarrow x_n}=C+\\
&\log\int_{\vx\backslash x_n,\boldsymbol\theta}\Big[\vphantom{\textstyle\sum_{j\neq n}\Delta^{(t)}_{x_j\rightarrow\Phi_m}}\Phi_m\left(y_m, \vx, \boldsymbol\theta\right)\cdot\exp\Big(\sum_{j\neq n}\Delta^{(t)}_{x_j\rightarrow\Phi_m}+\sum_v\Delta^{(t)}_{\theta_v\rightarrow\Phi_m}\Big)\Big]
\end{split}\\
\label{eq:sp_fv_omega_x}
&\Delta^{(t+1)}_{\Omega_n\rightarrow x_n}=C+\log\int_{\boldsymbol\lambda}\Omega_n(x_n,\boldsymbol\lambda)\cdot\exp\Big(\sum_u\Delta^{(t)}_{\lambda_u\rightarrow\Omega_n}\Big)\\
\label{eq:sp_vf_x_phi}
&\Delta^{(t+1)}_{x_n\rightarrow \Phi_m}=\Delta^{(t+1)}_{\Omega_n\rightarrow x_n}+\sum_{i\neq m}\Delta^{(t+1)}_{\Phi_i\rightarrow x_n}\\
\label{eq:sp_vf_x_omega}
&\Delta^{(t+1)}_{x_n\rightarrow \Omega_n}=\sum_i\Delta^{(t+1)}_{\Phi_i\rightarrow x_n}\,,
\end{align}
\end{subequations}
where $C$ (by abuse of notation\footnote{Note that the $C$ in \eqref{eq:sp_fv_phi_x} and the $C$ in \eqref{eq:sp_fv_omega_x} are in fact different, they are both some constants in the $(t+1)$-th iteration.}) denotes some constant that does not depend on the messages in the previous $t$-th iteration, $\vx\backslash x_n$ is the vector $\vx$ with its $n$-th entry $x_n$ removed, $\Phi_m(y_m,\vx,\boldsymbol\theta)=p(y_m|\vx,\boldsymbol\theta)$ and $\Omega_n(x_n,\boldsymbol\lambda)= p(x_n|\boldsymbol\lambda)$. 

In the \emph{signal prior} block, the messages exchanged among the nodes in the $(t+1)$-th iteration are:
\begin{subequations}
\label{eq:sp_fv_parameter}
\begin{align}
\label{eq:sp_fv_omega_lambda}
\begin{split}
&\Delta^{(t+1)}_{\Omega_n\rightarrow \lambda_l}=C+\\
&\log\int_{x_n,\boldsymbol\lambda\backslash\lambda_l}\Big[\Omega_n(x_n,\boldsymbol\lambda)\cdot\exp\Big(\Delta^{(t+1)}_{x_n\rightarrow \Omega_n}+\sum_{u\neq l}\Delta^{(t)}_{\lambda_u\rightarrow\Omega_n}\Big)\Big]
\end{split}\\
\label{eq:sp_vf_lambda_omega}
&\Delta^{(t+1)}_{\lambda_l\rightarrow\Omega_n}=C+\sum_{j\neq n}\Delta^{(t+1)}_{\Omega_j\rightarrow\lambda_l}\,,
\end{align}
\end{subequations}
where $\boldsymbol\lambda\backslash\lambda_l$ is the set $\boldsymbol\lambda$ with its element $\lambda_l$ removed.

In the \emph{noise prior} block, we have the following messages in the $(t+1)$-th iteration:
\begin{subequations}
\begin{align}
\label{eq:sp_fv_phi_theta}
\begin{split}
&\Delta^{(t+1)}_{\Phi_m\rightarrow\theta_k}=C+\\
&\log\int_{\boldsymbol\theta\backslash\theta_k,\vx}\Big[\Phi_m\left(y_m, \vx, \boldsymbol\theta\right)\cdot\exp\Big(\sum_j\Delta^{(t)}_{x_j\rightarrow\Phi_m}+\sum_{v\neq k}\Delta^{(t)}_{\theta_v\rightarrow\Phi_m}\Big)\Big]
\end{split}\\
\label{eq:sp_vf_theta_phi}
&\Delta^{(t+1)}_{\theta_k\rightarrow\Phi_m}=\sum_{i\neq m}\Delta^{(t+1)}_{\Phi_i\rightarrow\theta_k}\,,
\end{align}
\end{subequations}
where $\boldsymbol\theta\backslash\theta_k$ is the set $\boldsymbol\theta$ with its element $\theta_k$ removed. 

The posteriors of the signal $\vx$ and the distribution parameters $\boldsymbol\lambda,\boldsymbol\theta$ are then:

\begin{subequations}
\label{eq:sm_post_dist}
\begin{align}
\label{eq:pm_x}
\begin{split}
p(x_n|\vy)&\propto\exp\Big(\Delta^{(t+1)}_{\Omega_n\rightarrow x_n}+\sum_m\Delta^{(t+1)}_{\Phi_m\rightarrow x_n}\Big)
\end{split}\\
\label{eq:pm_lambda}
\begin{split}
p(\lambda_l|\vy)&\propto\exp\Big(\sum_n\Delta^{(t+1)}_{\Omega_n\rightarrow\lambda_l}\Big)
\end{split}\\
\label{eq:pm_theta}
\begin{split}
p(\theta_k|\vy)&\propto\exp\Big(\sum_m\Delta^{(t+1)}_{\Phi_m\rightarrow\theta_k}\Big)\,.
\end{split}
\end{align}
\end{subequations}

\subsection{Parameter Estimation}
\label{sec:pe}
The distribution parameters $\boldsymbol\lambda,\boldsymbol\theta$ can be estimated by maximizing the posteriors in \eqref{eq:pm_lambda} and \eqref{eq:pm_theta}.
\begin{subequations}
\begin{align}
    \label{eq:lambda_est}
    \hat{\lambda}_l^{(t+1)} &= \arg\max_{\lambda_l}p(\lambda_l|\vy)=\arg\max_{\lambda_l}\sum_n\Delta^{(t+1)}_{\Omega_n\rightarrow\lambda_l}\\
    \label{eq:theta_est}
    \hat{\theta}_k^{(t+1)} &= \arg\max_{\theta_k}p(\theta_k|\vy)=\arg\max_{\theta_k}\sum_m\Delta^{(t+1)}_{\Phi_m\rightarrow\theta_k}\,.
\end{align}
\end{subequations}
We shall combine EM and the second-order method to find the maximizing parameters in \eqref{eq:lambda_est}, \eqref{eq:theta_est}, which turns out to be a much simpler alternative to previous EM-based approaches that maximize the measurement likelihood \cite{Vila:EMGM:2013,Kamilov:PE:2014}. As discussed later in Section \ref{subsec:comparison_with_pre}, this subtle modification allows us to consider the much more complicated quantization noise models that often arise from applications such as channel estimation in the massive MIMO systems.

Using the estimated parameters $\hat{\boldsymbol\lambda},\hat{\boldsymbol\theta}$, we can simplify the messages passed from the factor nodes to the variable nodes as follows:
\begin{subequations}
\label{eq:sp_simplified_messages}
\begin{align}
\label{eq:delta_phi_x_simplifed}
\begin{split}
&\Delta^{(t+1)}_{\Phi_m\rightarrow x_n}=C+\\
&\log\int_{\vx\backslash x_n}\Big[\Phi_m\Big(y_m, \vx, \hat{\boldsymbol\theta}^{(t)}\Big)\cdot\exp\Big(\sum_{j\neq n}\Delta^{(t)}_{x_j\rightarrow\Phi_m}\Big)\Big]
\end{split}\\
\label{eq:delta_omega_x_simplified}
&\Delta^{(t+1)}_{\Omega_n\rightarrow x_n}=C+\log\Omega_n\Big(x_n,\hat{\boldsymbol\lambda}^{(t)}\Big)\\
\label{eq:delta_omega_lambda_simplified}
\begin{split}
&\Delta^{(t+1)}_{\Omega_n\rightarrow \lambda_l}=C+\log\int_{x_n}\left[\Omega_n\left(x_n,\lambda_l,\hat{\boldsymbol\lambda}^{(t)}\backslash\hat{\lambda}_l^{(t)}\right)\cdot\exp\left(\Delta^{(t+1)}_{x_n\rightarrow \Omega_n}\right)\right]
\end{split}\\
\label{eq:delta_phi_theta_simplified}
\begin{split}
&\Delta^{(t+1)}_{\Phi_m\rightarrow\theta_k}=C+\\
&\log\int_{\vx}\Big[\Phi_m\Big(y_m, \vx, \theta_k, \hat{\boldsymbol\theta}^{(t)}\backslash\hat{\theta}_k^{(t)}\Big)\cdot\exp\Big(\sum_j\Delta^{(t)}_{x_j\rightarrow\Phi_m})\Big]\,,
\end{split}
\end{align}
\end{subequations}
where $\hat{\boldsymbol\lambda}^{(t)}$, $\hat{\boldsymbol\theta}^{(t)}$ are estimated parameters from the previous $t$-th iteration. 

\section{1-Bit Compressive Sensing via AMP}
\label{sec:1bitCS}
In this section we introduce the sparse signal model and 1-bit quantization noise model under the Bayesian setting, and show how the signal and the parameters can be jointly recovered via AMP. 
\begin{sparse_signal_model}
The entries of the sparse signal $\vx$ are assumed to be i.i.d.
\begin{align}
    p(\vx|\boldsymbol\lambda)=\prod_n p(x_n|\boldsymbol\lambda)\,.
\end{align}
The Bernoulli and Gaussian mixture distribution is used to model the sparse signal $x_n$
\begin{align}
\label{eq:bgm}
    p(x_n|\boldsymbol\lambda) = (1-\kappa)\cdot\delta(x_n)+\kappa\cdot\sum_i\xi_i\cdot\mathcal{N}(x_n|\mu_i,{\gamma_x}_i)\,,
\end{align}
where $\delta(x_n)$ is the Dirac delta function, $\kappa$ is the probability that $x_n$ takes a non-zero value, $\xi_i$ is the Gaussian mixture weights, $\mu_i$ and ${\gamma_x}_i$ are the mean and variance of the $i$-th Gaussian component, $\mathcal{N}(x_n|\mu_i,{\gamma_x}_i)$ is the Gaussian probability density function. In this case the parameter set $\boldsymbol\lambda$ is
\begin{align}
\label{eq:parameter_set_lambda}
\boldsymbol\lambda=\left\{\left.\kappa,\xi_i,\mu_i,{\gamma_x}_i\ \right|\ i=1,\cdots,D\right\}\,,
\end{align}
where $D$ is the number of Gaussian mixture components.
\end{sparse_signal_model}

\begin{figure}[tbp]
\centering
\includegraphics[width=0.3\textwidth]{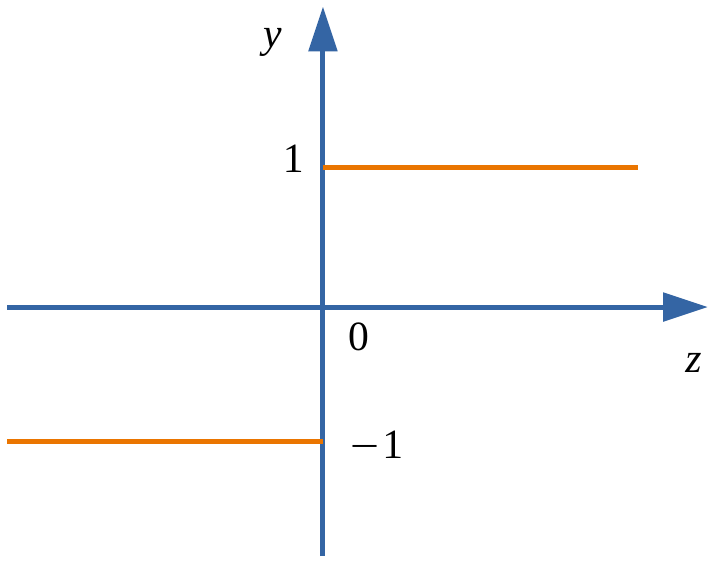}
\caption{The quantizer $\mathcal{Q}$ outputs the sign of the input $z$ in 1-bit compressive sensing.}
\label{fig:1bit_quantizer}
\end{figure}

\begin{1_bit_quantization_noise_model}
The noisy 1-bit measurements $\vy$ are
\begin{align}
\vy=\mathscr{Q}(\mA\vx+\vw)\,, \tag{\ref{eq:quantized_measurements} revisited}
\end{align}
where $\vw$ is the i.i.d. additive white Gaussian noise (AWGN) with $w_m\sim\mathcal{N}(0,\gamma_w)$, $\gamma_w$ is the noise variance. $\vw$ is added to the noiseless measurements $\vz=\mA\vx$ before quantization. As shown in Fig. \ref{fig:1bit_quantizer}, the element-wise quantizer $\mathscr{Q}$ produces 1-bit output by computing the sign of an input $z_m+w_m$
\begin{align}
\label{eq:1bit_quantization}
    y_m=\mathscr{Q}(z_m+w_m)=\left\{
    \begin{array}{l}
    +1  \\
    -1 
    \end{array}
    \quad
    \begin{array}{l}
    \textrm{if }z_m+w_m>0\\
    \textrm{if }z_m+w_m\leq0\,.
    \end{array}
    \right.
\end{align}
In this case the parameter set $\boldsymbol\theta$ is simplify
\begin{align}
\label{eq:parameter_set_theta}
    \boldsymbol\theta=\left\{\gamma_w\right\}\,.
\end{align}
\end{1_bit_quantization_noise_model}

We use the GAMP formulation in \cite{Rangan:GAMP:2011} to compute the messages $\Delta_{\Phi_m\rightarrow x_n}$ and $\Delta_{x_n\rightarrow\Phi_m}$. The proposed AMP with built-in parameter estimation (AMP-PE) algorithm is summarized in Algorithm \ref{alg:amp_pe}. The damping and mean removal operations are often incorporated to the AMP algorithm to ensure convergence for ill-conditioned or non-zero-mean measurement matrix \cite{Rangan:DampingCvg:2014,Vila:DampingMR:2015}. Note that in \eqref{eq:compute_s_m},\eqref{eq:compute_tau_s_m},\eqref{eq:compute_x_n},\eqref{eq:compute_tau_x_n}, we need to compute the posterior means and variances of $\vz$ and $\vx$. For the sparse signal model and 1-bit quantization noise model introduced here, their expressions can be derived as follows.

\begin{algorithm}[tbp]
\caption{The AMP-PE algorithm }\label{alg:amp_pe}
\begin{algorithmic}[1]
\Require $\hat{\vx}^{(0)}, \boldsymbol\tau_x^{(0)}, \vq^{(0)}, \boldsymbol\tau_q^{(0)}, \hat{\boldsymbol\lambda}^{(0)}, \hat{\boldsymbol\theta}^{(0)}$.

\For{$t=\{0,1,\cdots,T\}$}
	\State Output \emph{nonlinear} update: For each $m=1,\cdots,M$
	\begin{subequations}
	\begin{align}
	\label{eq:compute_s_m}
	s_m^{(t)}&=\frac{1}{{\tau_q}^{(t)}}\left(\mathbb{E}\left[z_m\left|q_m^{(t)}, {\tau_q}^{(t)}, y_m, \hat{\boldsymbol\theta}^{(t)}\right.\right]-q_m^{(t)}\right)\\
	\label{eq:compute_tau_s_m}
	{\tau_s}^{(t)}&=\frac{1}{M}\sum_m\frac{1}{{\tau_q}^{(t)}}\left(1-\frac{1}{{\tau_q}^{(t)}}\textrm{Var}\left[z_m\left|q_m^{(t)}, {\tau_q}^{(t)}, y_m, \hat{\boldsymbol\theta}^{(t)}\right.\right]\right).
	\end{align}
	\end{subequations}
	\State Input \emph{linear} update: For each $n=1,\cdots,N$
	\begin{subequations}
	\begin{align}
	{\tau_r}^{(t)}&=\left[\frac{1}{N}\|\mA\|_F^2\cdot {\tau_s}^{(t)}\right]^{-1}\\
	r_n^{(t)}&=x_n^{(t)}+{\tau_r}^{(t)}\sum_mA_{mn}\cdot s_m^{(t)}\,.
	\end{align}
	\end{subequations}
	\State Estimate the input parameters: For each $l=1,\cdots,L$
	\begin{align}
	    \label{eq:amp_pe_est_input}
	    \hat{\lambda}_l^{(t+1)}&=\arg\max_{\lambda_l}\ \sum_n\Delta_{\Omega_n\rightarrow\lambda_l}^{(t+1)}\,.
	\end{align}
	\State Input \emph{nonlinear} update: For each $n=1,\cdots,N$
	\begin{subequations}
	\begin{align}
	\label{eq:compute_x_n}
	\hat{x}_n^{(t+1)} &= \mathbb{E}\left[x_n\left|r_n^{(t)}, {\tau_r}^{(t)}, \hat{\boldsymbol\lambda}^{(t+1)}\right.\right]\\
	\label{eq:compute_tau_x_n}
	{\tau_x}^{(t+1)}&=\frac{1}{N}\sum_n\textrm{Var}\left[x_n\left|r_n^{(t)}, {\tau_r}^{(t)}, \hat{\boldsymbol\lambda}^{(t+1)}\right.\right]\,.
	\end{align}
	\end{subequations}
	\State Output \emph{linear} update: For each $m=1,\cdots,M$
	\begin{subequations}
	\begin{align}
	{\tau_q}^{(t+1)}&=\frac{1}{M}\|\mA\|_F^2\cdot {\tau_x}^{(t+1)}\\
	q_m^{(t+1)}&=\sum_nA_{mn}\cdot \hat{x}_n^{(t+1)}-{\tau_q}^{(t+1)}\cdot s_m^{(t)}\,.
	\end{align}
	\end{subequations}
	\State Estimate the output parameters: For each $k=1,\cdots,K$
	\begin{align}
	    \label{eq:amp_pe_est_output}
	    \hat{\theta}_k^{(t+1)}&=\arg\max_{\theta_k}\ \sum_m\Delta_{\Phi_m\rightarrow\theta_k}^{(t+1)}\,.
	\end{align}
	\If {$\hat{\vx}^{(t+1)}$ reaches convergence}
		\State $\hat{\vx}=\hat{\vx}^{(t+1)}$ and \textbf{break};
	\EndIf
\EndFor
\State\Return The recovered signal $\hat{\vx}$;
\end{algorithmic}
\end{algorithm}

\subsection{Nonlinear Updates for the Sparse Signal Model}
The Bernoulli and Gaussian mixture model is chosen as the sparse signal model in this paper. To simplify the notations, we remove the superscript that denotes the iteration index $t$ in the following derivations. Under the sum-product message passing, the posterior of the signal $\vx$ is approximated as
\begin{align}
    p(x_n|\vy)\approx\frac{1}{\Psi(r_n)} p(x_n|\boldsymbol\lambda)\cdot\mathcal{N}(x_n|r_n,{\tau_r})\,,
\end{align}
where $\Psi(r_n)$ is the normalizing constant
\begin{align}
\begin{split}
    &\Psi(r_n)=\int_{x_n} p(x_n|\boldsymbol\lambda)\cdot\mathcal{N}(x_n|r_n,{\tau_r})\\
    &=(1-\kappa)\mathcal{N}(r_n|0,{\tau_r})+\sum_i\kappa\xi_i\cdot\mathcal{N}(r_n|\mu_i,{\gamma_x}_i+{\tau_r})\,.
\end{split}
\end{align}
The posterior mean of $x_n$ in \eqref{eq:compute_x_n} can then be computed as
\begin{align}
\begin{split}
    \hat{x}_n&=\mathbb{E}\left[x_n|r_n,{\tau_r},\boldsymbol\lambda\right]\\
    &=\frac{1}{\Psi(r_n)}\sum_i\kappa\xi_i\cdot\mathcal{N}(r_n|\mu_i,{\gamma_x}_i+{\tau_r})\frac{\mu_i{\tau_r}+r_n{\gamma_x}_i}{{\gamma_x}_i+{\tau_r}}\,.
\end{split}
\end{align}
The posterior expectation of $x_n^2$ is
\begin{align}
\begin{split}
\mathbb{E}\left[x_n^2|r_n,{\tau_r},\boldsymbol\lambda\right] =& \frac{1}{\Psi(r_n)}\sum_i\kappa\xi_i\cdot\mathcal{N}(r_n|\mu_i,{\gamma_x}_i+{\tau_r})\\
&\times\left(\frac{{\gamma_x}_i{\tau_r}}{{\gamma_x}_i+{\tau_r}}+\left(\frac{\mu_i{\tau_r}+r_n{\gamma_x}_i}{{\gamma_x}_i+{\tau_r}}\right)^2\right)\,.
\end{split}
\end{align}
The posterior variance of $x_n$ in \eqref{eq:compute_tau_x_n} is then
\begin{align}
    {\tau_x}=\mathbb{E}\left[x_n^2|r_n,{\tau_r},\boldsymbol\lambda\right]-\left(\mathbb{E}\left[x_n|r_n,{\tau_r},\boldsymbol\lambda\right]\right)^2\,.
\end{align}

\subsection{Nonlinear Updates for the 1-Bit Quantization Noise Model}
From \eqref{eq:1bit_quantization}, we can get that
\begin{align}
    \label{eq:prob_1bit_1}
    \mathrm{Pr}\left(y_m=1|z_m\right)&=\int_{-\infty}^0\mathcal{N}(u|z_m,\gamma_w)\ du\\
    \label{eq:prob_1bit_2}
    \mathrm{Pr}\left(y_m=-1|z_m\right)&=1-\int_{-\infty}^0\mathcal{N}(u|z_m,\gamma_w)\ du\,.
\end{align}
To simplify the notations, we define the following
\begin{align}
\label{eq:q_bar}
&\overline{q}_m\coloneqq\frac{q_m}{\sqrt{{\tau_q}+\gamma_w}}\\
\label{eq:h0_q_bar}
&h_0(q_m)\coloneqq\int_\infty^0\mathcal{N}(u|q_m,{\tau_q}+\gamma_w)\ du=\frac{1}{2}\textrm{erfc}\left(\sqrt{\frac{1}{2}}\cdot\overline{q}_m\right)\\
\label{eq:U0_q_bar_ym}
&\mathcal{U}_0(q_m,y_m)\coloneqq\big(1-h_0(q_m)\big)\delta(y_m-1)+ h_0(q_m)\delta(y_m+1)\,,
\end{align}
where $\textrm{erfc}(\cdot)$ is the ``complementary error function''.

According to the sum-product message passing, the posterior of $z_m$ can be approximated as
\begin{align}
    p(z_m|y_m)\approx\frac{1}{\mathcal{U}_0(q_m,y_m)}p(y_m|z_m)\cdot\mathcal{N}(z_m|q_m,{\tau_q})\,.
\end{align}
where $\mathcal{U}_0(q_m,y_m)$ is the normalizing constant
\begin{align}
\begin{split}
    \mathcal{U}_0(q_m,y_m)&=\int_{z_m}p(y_m|z_m)\cdot\mathcal{N}(z_m|q_m,{\tau_q})\,.
\end{split}
\end{align}

We further define the following
\begin{align}
    \begin{split}
        h_1(q_m)\coloneqq&\int z_m 
        \int_{-\infty}^0\mathcal{N}(u|z_m,\gamma_w)\ du\cdot\mathcal{N}(z_m|q_m,{\tau_q})\ dz_m\\
        =&q_m\cdot h_0(q_m)-{\tau_q}\cdot\mathcal{N}(q_m|0,{\tau_q}+\gamma_w)
    \end{split}\\
    \begin{split}
        h_2(q_m)\coloneqq&\int z_m^2\int_{-\infty}^0\mathcal{N}(u|z_m,\gamma_w)\ du\cdot\mathcal{N}(z_m|q_m,{\tau_q})\ dz_m\\
        =&(q_m^2+{\tau_q})\cdot h_0(q_m)\\
        &-q_m\cdot\frac{{\tau_q}^2+2{\tau_q}\gamma_w}{{\tau_q}+\gamma_w}\cdot\mathcal{N}(q_m|0,{\tau_q}+\gamma_w)
    \end{split}
\end{align}
\vspace{-1em}
\begin{align}
    \mathcal{U}_1(q_m,y_m)\coloneqq&\big(q_m-h_1(q_m)\big)\delta(y_m-1)+ h_1(q_m)\delta(y_m+1)\\
    \begin{split}
    \mathcal{U}_2(q_m,y_m)\coloneqq&\big(q_m^2+{\tau_q}-h_2(q_m)\big)\delta(y_m-1)\\
    &+ h_2(q_m)\delta(y_m+1)\,.
    \end{split}
\end{align}
The posterior mean of $z_m$ in \eqref{eq:compute_s_m} can be computed as
\begin{align}
\begin{split}
    \mathbb{E}\left[z_m\left|q_m, {\tau_q}, y_m, \boldsymbol\theta\right.\right]&=\int_{z_m}z_m\cdot p(z_m|y_m)\\
    &=\frac{1}{\mathcal{U}_0(q_m,y_m)}\mathcal{U}_1(q_m,y_m)\,.
\end{split}
\end{align}
The posterior expectation of $z_m^2$ is 
\begin{align}
\begin{split}
    \mathbb{E}\left[z_m^2\left|q_m, {\tau_q}, y_m, \boldsymbol\theta\right.\right]&=\int_{z_m}z_m^2\cdot p(z_m|y_m)\\
    &=\frac{1}{\mathcal{U}_0(q_m,y_m)}\mathcal{U}_2(q_m,y_m)\,.
\end{split}
\end{align}
The posterior variance of $z_m$ in \eqref{eq:compute_tau_s_m} is then
\begin{align}
\begin{split}
    {\tau_z} &= \mathbb{E}\left[z_m^2\left|q_m, {\tau_q}, y_m, \boldsymbol\theta\right.\right]-\left(\mathbb{E}\left[z_m\left|q_m, {\tau_q}, y_m, \boldsymbol\theta\right.\right]\right)^2\\
    &=\frac{1}{\mathcal{U}_0(q_m,y_m)}\big(\mathcal{U}_1(q_m,y_m)-\mathcal{U}_2(q_m,y_m)\big)\,.
\end{split}
\end{align}

\subsection{Parameter Estimation for the Sparse Signal Model}
\label{subsec:pe_sparse_signal_model}

We first show how to estimate the signal prior parameters $\boldsymbol\lambda$. Combining \eqref{eq:lambda_est}, \eqref{eq:delta_omega_lambda_simplified}, \eqref{eq:bgm} and \eqref{eq:parameter_set_lambda}, we have
\begin{align}
\label{eq:pe_lambda_signal_prior}
\begin{split}
    \hat{\boldsymbol\lambda} = \arg\max_{\boldsymbol\lambda}\sum_n\log\Big[&(1-\kappa)\cdot\mathcal{N}(r_n|0,{\tau_r})\\
    &+\sum_i\kappa\xi_i\cdot\mathcal{N}(r_n|\mu_i,{\gamma_x}_i+{\tau_r})\Big]\,.
\end{split}
\end{align}
Standard gradient descent was previously used to solve the above \eqref{eq:pe_lambda_signal_prior} in \cite{PE_GAMP17}. However, incremental searches along the gradient direction significantly slow down the algorithm -- a major disadvantage when the problem size is large. In this paper, we shall rely on EM to estimate the signal prior parameters $\boldsymbol\lambda$, and then switch to the second-order method to estimate the noise prior parameters $\boldsymbol\theta$. The second-order method computes the search step size adaptively based on the current solution, which is more computationally efficient.

Here $r_n$ is treated as the observation, and the latent variable $c(r_n)\in\{0,1,\cdots,D\}$ determines which mixture component $r_n$ comes from. In the $(e+1)$-th EM iteration, letting $f(\boldsymbol\lambda)$ be the objective function computed from the expectation step, we maximize $f(\boldsymbol\lambda)$ in the maximization step
\begin{align}
\begin{split}
    \hat{\boldsymbol\lambda}^{(e+1)}=&\arg\max_{\boldsymbol\lambda} f(\boldsymbol\lambda)\\
    =&\arg\max_{\boldsymbol\lambda}\sum_n\psi_0(r_n)\cdot\log\left[(1-\kappa)\cdot\mathcal{N}(r_n|0,{\tau_r})\right]\\
    &+\sum_n\sum_i\psi_i(r_n)\cdot\log\left[\kappa\xi_i\cdot\mathcal{N}(r_n|\mu_i,{\gamma_x}_i+{\tau_r})\right]\,,
\end{split}
\end{align}
where $\psi_0(r_n)$ and $\psi_i(r_n)$ are the posteriors of the latent variable $c(r_n)$:
\begin{align}
    \psi_0(r_n)&=\frac{1}{\Psi(r_n)}\big(1-\kappa^{(e)}\big)\cdot\mathcal{N}(r_n|0,{\tau_r})\\
    \psi_i(r_n)&=\frac{1}{\Psi(r_n)}\kappa^{(e)}\xi_i^{(e)}\cdot\mathcal{N}\big(r_n|\mu_i^{(e)},{\gamma_x}_i^{(e)}+{\tau_r}\big)\\
    \begin{split}
    \Psi(r_n)&=\big(1-\kappa^{(e)}\big)\cdot\mathcal{N}(r_n|0,{\tau_r})\\
    &\quad+\sum_i\kappa^{(e)}\xi_i^{(e)}\cdot\mathcal{N}\big(r_n|\mu_i^{(e)},{\gamma_x}_i^{(e)}+{\tau_r}\big)\,.
    \end{split}
\end{align}
The mixture weights $\kappa,\xi_i$ and the Gaussian mixture mean $\mu_i$ can be updated as follows
\begin{align}
    \kappa^{(e+1)} &= \frac{\sum_n\sum_i\psi_i(r_n)}{\sum_n\psi_0(r_n)+\sum_n\sum_i\psi_i(r_n)}\\
    \xi_i^{(e+1)} &= \frac{\sum_n\psi_i(r_n)}{\sum_n\sum_i\psi_i(r_n)}\\
    \mu_i^{(e+1)} &=\frac{\sum_n\psi_i(r_n)\cdot\frac{r_n}{{\gamma_x}_i^{(e)}+{\tau_r}}}{\sum_n\psi_i(r_n)\cdot\frac{1}{{\gamma_x}_i^{(e)}+{\tau_r}}}\,.
\end{align}

Although we could not obtain a closed-form update for the Gaussian mixture variance ${\gamma_x}_i$ that maximizes $f(\boldsymbol\lambda)$, we can maximize the second order approximation of $f(\boldsymbol\lambda)$ at $\boldsymbol\lambda^{(e)}$ instead.
\begin{align}
    f(\boldsymbol\lambda)\approx f(\boldsymbol\lambda^{(e)})+f^\prime\cdot\big(\boldsymbol\lambda-\boldsymbol\lambda^{(e)}\big)+\frac{f^{\prime\prime}}{2}\cdot\big(\boldsymbol\lambda-\boldsymbol\lambda^{(e)}\big)^2\,.
\end{align}
The first order and second order derivatives of $f(\boldsymbol\lambda)$ with respect to $\boldsymbol\lambda$ are
\begin{align}
    f^\prime({\gamma_x}_i)&=\sum_n\psi_i(r_n)\left[\frac{1}{2}\cdot\frac{\big(r_n-\mu_i^{(e)}\big)^2}{({\tau_r}+{\gamma_x}_i)^2}-\frac{1}{2}\cdot\frac{1}{{\tau_r}+{\gamma_x}_i}\right]\\
    f^{\prime\prime}({\gamma_x}_i)&=\sum_n\psi_i(r_n)\left[-\frac{\big(r_n-\mu_i^{(e)}\big)^2}{({\tau_r}+{\gamma_x}_i)^3}+\frac{1}{2}\cdot\frac{1}{({\tau_r}+{\gamma_x}_i)^2}\right].
\end{align}
Note that the second order method does not always give us the maximizing solution. We should use use gradient descent to compute it when $f^{\prime\prime}\geq0$. When $f^{\prime\prime}<0$, the update for ${\gamma_x}_i$ is then
\begin{align}
    \label{eq:gm_var_update}
    {\gamma_x}_i^{(e+1)} &= {\gamma_x}_i^{(e)}-\frac{f^\prime\big({\gamma_x}_i^{(e)}\big)}{f^{\prime\prime}\big({\gamma_x}_i^{(e)}\big)}\,.
\end{align}

\subsection{Parameter Estimation for the 1-Bit Quantization Noise Model}
We next show how to estimate the noise prior parameters. Combining \eqref{eq:theta_est}, \eqref{eq:delta_phi_theta_simplified}, \eqref{eq:1bit_quantization} and \eqref{eq:parameter_set_theta}, we have
\begin{align}
\label{eq:g1_theta}
\begin{split}
    \hat{\boldsymbol\theta}=\arg\max_{\boldsymbol\theta}g_1(\boldsymbol\theta)=\arg\max_{\boldsymbol\theta}\sum_m\log\left[\mathcal{U}_0(q_m,y_m)\right]\,.
\end{split}
\end{align}
We also could not obtain a closed form update for the AWGN variance $\gamma_w$ that maximizes $g_1(\boldsymbol\theta)$. Here we maximize the second order approximation of $g_1(\boldsymbol\theta)$ at $\boldsymbol\theta^{(e)}$ instead.
\begin{align}
    g_1(\boldsymbol\theta)\approx g_1(\boldsymbol\theta^{(e)})+g_1^\prime\cdot\big(\boldsymbol\theta-\boldsymbol\theta^{(e)}\big)+\frac{g_1^{\prime\prime}}{2}\cdot\big(\boldsymbol\theta-\boldsymbol\theta^{(e)}\big)^2\,.
\end{align}
The first and second orders derivatives of $g_1(\boldsymbol\theta)$ with respect to $\gamma_w$ are 
\begin{align}
    g_1^\prime(\gamma_w) &= \sum_m\frac{1}{\mathcal{U}_0(q_m,y_m)}\cdot\frac{\partial\mathcal{U}_0}{\partial\gamma_w}\\
    \begin{split}
    g_1^{\prime\prime}(\gamma_w) &= \sum_m-\frac{1}{\left(\mathcal{U}_0(q_m,y_m)\right)^2}\left[\frac{\partial\mathcal{U}_0}{\partial\gamma_w}\right]^2+\frac{1}{\mathcal{U}_0(q_m,y_m)}\cdot\frac{\partial^2\mathcal{U}_0}{\partial\gamma_w^2}\,,
    \end{split}
\end{align}
where the first and second order derivatives of $\mathcal{U}_0$ are
\begin{align}
    \frac{\partial\mathcal{U}_0}{\partial\gamma_w}&=-\frac{\partial h_0}{\partial\gamma_w}\cdot\delta(y_m-1)+\frac{\partial h_0}{\partial\gamma_w}\cdot\delta(y_m+1)\\
    \frac{\partial^2\mathcal{U}_0}{\partial\gamma_w^2}&=-\frac{\partial^2 h_0}{\partial\gamma_w^2}\cdot\delta(y_m-1)+\frac{\partial^2 h_0}{\partial\gamma_w^2}\cdot\delta(y_m+1)\,,
\end{align}
with the first and second order derivatives of $h_0(q_m)$ with respect to $\gamma_w$ given by
\begin{align}
    \frac{\partial h_0}{\partial\gamma_w}=&\frac{\overline{q}_m}{2\sqrt{2\pi}(\gamma_w+{\tau_q})}\cdot\exp\left(-\frac{\overline{q}_m^2}{2}\right)\\
    \frac{\partial^2h_0}{\partial\gamma_w^2}=&\frac{\overline{q}_m^3-3\overline{q}_m}{4\sqrt{2\pi}(\gamma_w+{\tau_q})^2}\cdot\exp\left(-\frac{\overline{q}_m^2}{2}\right)\,.
\end{align}

Similarly, when $g_1^{\prime\prime}\geq0$, gradient descent is used. When $g_1^{\prime\prime}<0$, the update for $\gamma_w$ is then
\begin{align}
\label{eq:1bit_awgn_variance_update}
    \gamma_w^{(e+1)}=\gamma_w^{(e)}-\frac{g_1^\prime\big(\gamma_w^{(e)}\big)}{g_1^{\prime\prime}\big(\gamma_w^{(e)}\big)}\,.
\end{align}

\subsection{Comparison with Previous EM-based Approaches}
\label{subsec:comparison_with_pre}
In this subsection, we discuss the differences between the proposed AMP-PE approach and previous approaches that also utilize an EM-style strategy to estimate the distribution parameters. First, AMP-PE only uses EM to estimate the signal prior parameters, it then switches to the second-order method to estimate the noise prior parameter. Whereas previous approaches use EM to estimate \emph{both} the signal prior and noise prior parameters \cite{Vila:BG:2011,Vila:EMGM:2013,Krzakala:2012:1}. Second, AMP-PE maximizes the posteriors of parameters in a different fashion. In the \emph{signal prior} block, the dummy variable $r_n$ is treated as the observation, and its mixture label $c(r_n)$ is treated as the hidden variable. This leads to the objective function $f(\boldsymbol\lambda)$ in \eqref{eq:pe_lambda_signal_prior}. In the \emph{noise prior} block, the objective function is $g_1(\boldsymbol\theta)$ in \eqref{eq:g1_theta}. Previous approaches essentially try to maximize the likelihood of measurements. The noisy measurement $y_m$ is treated as the observation, the signal $x_n$ and the noiseless measurement $z_m$ are treated as the hidden variables. This leads to the following drastically different objective functions of previous approaches:
\begin{align}
    \label{eq:previous_em_signal}
    \max_{\boldsymbol\lambda}&\ \sum_n\int p\big(x_n|\vy,\boldsymbol\lambda^{(t)}\big)\log p(\vy,x_n|\boldsymbol\lambda)\ dx_n\,,\\
    \label{eq:previous_em_noise}
    \max_{\boldsymbol\theta}&\ \sum_m\int p\big(z_m|y_m,\boldsymbol\theta^{(t)}\big)\log p(y_m,z_m|\boldsymbol\theta)\ dz_m\,.
\end{align}

For the BGM model in \eqref{eq:bgm}, the optimization problem in \eqref{eq:previous_em_signal} can be solved easily with closed-form solutions. However, when it comes to the quantization noise model in \eqref{eq:1bit_quantization}, the problem in \eqref{eq:previous_em_noise} does not have closed-form solutions. Its objective function is much more complicated and computationally prohibitive compared to the one from AMP-PE in \eqref{eq:g1_theta}. 

\section{Experimental Results}
\label{sec:exp}

\begin{figure*}[htbp]
\centering
\subfigure{
\includegraphics[width=0.3\textwidth]{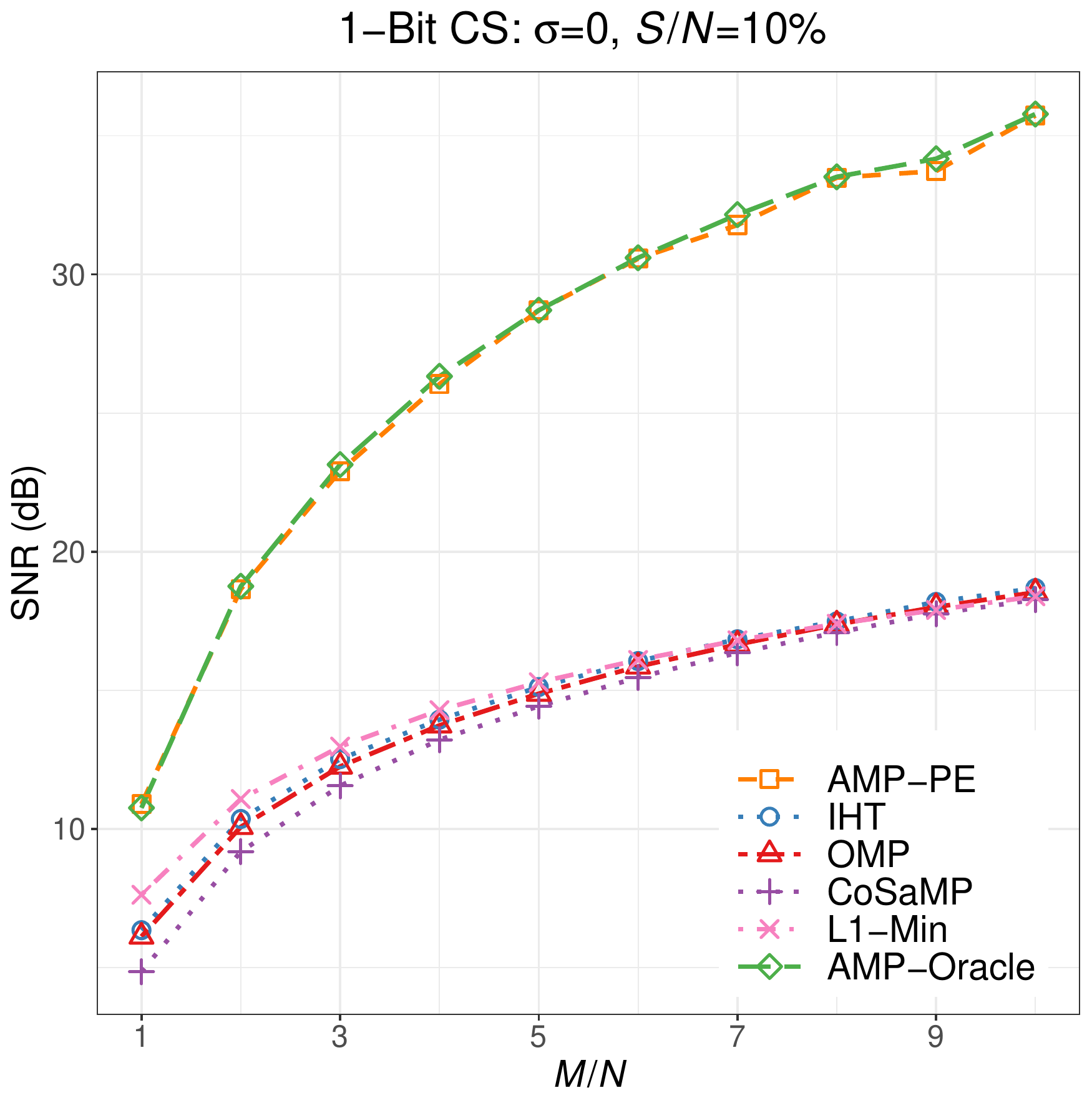}}
\subfigure{
\includegraphics[width=0.3\textwidth]{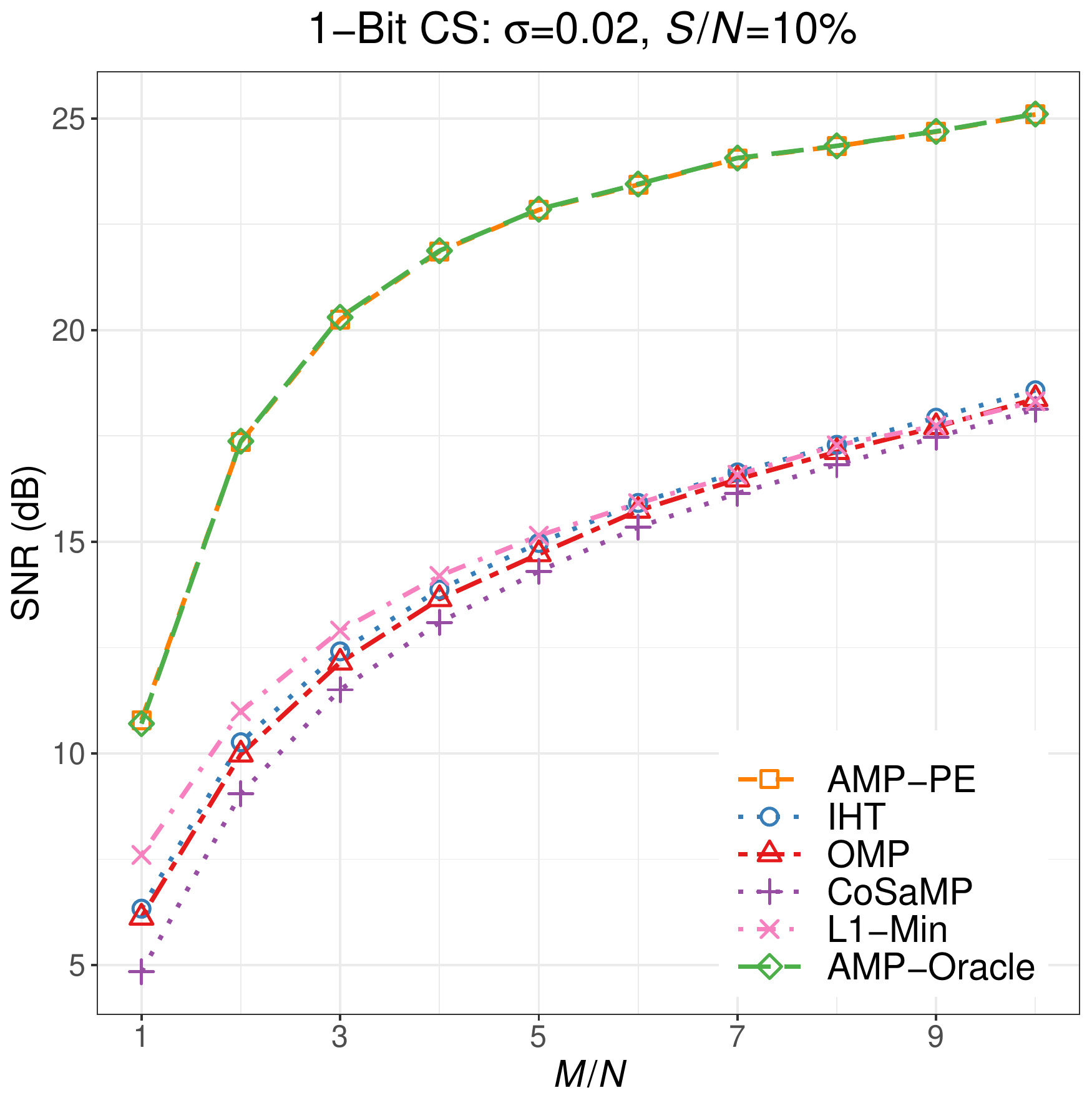}}
\subfigure{
\includegraphics[width=0.3\textwidth]{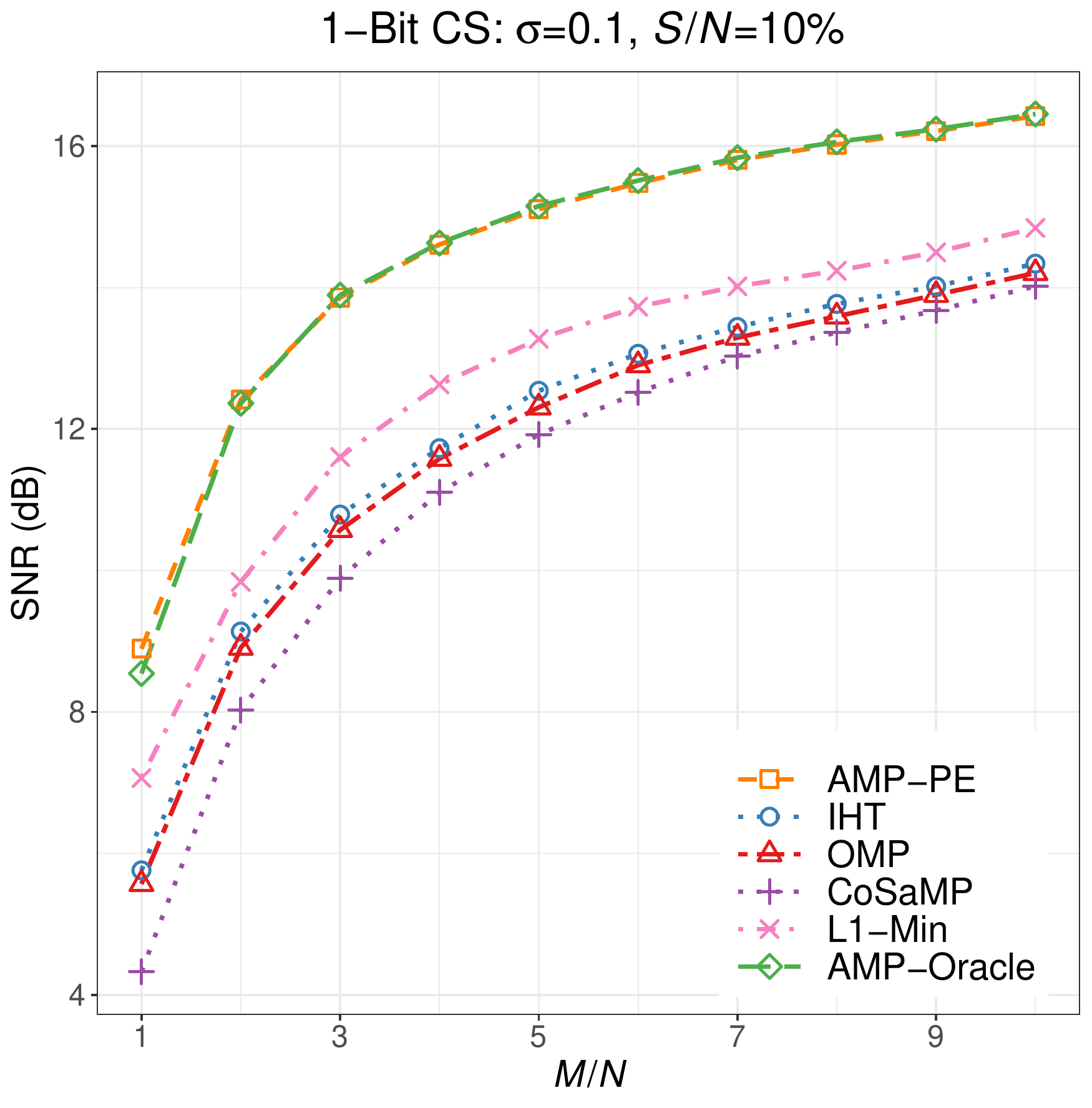}}\\
\subfigure{
\includegraphics[width=0.3\textwidth]{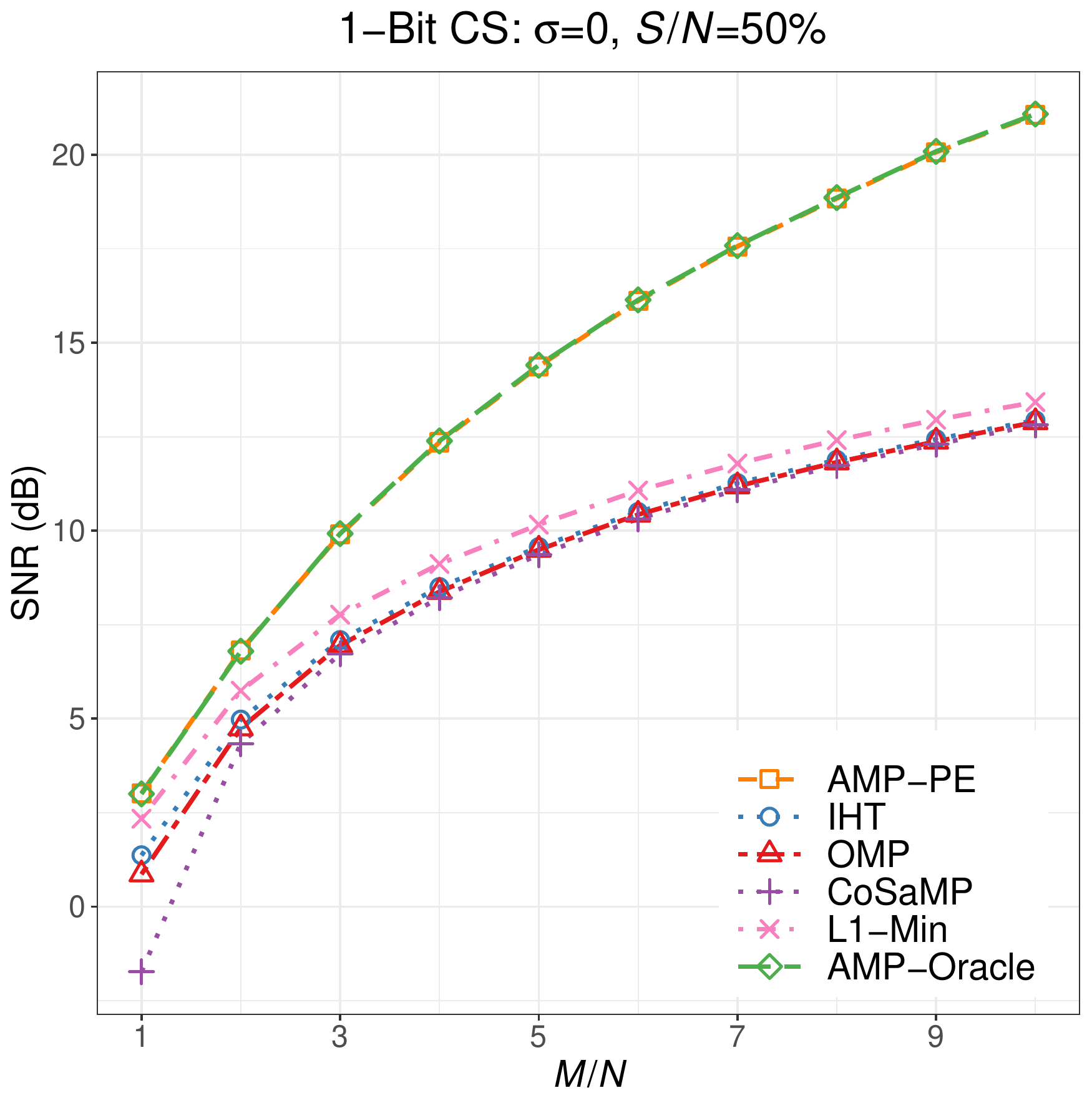}}
\subfigure{
\includegraphics[width=0.3\textwidth]{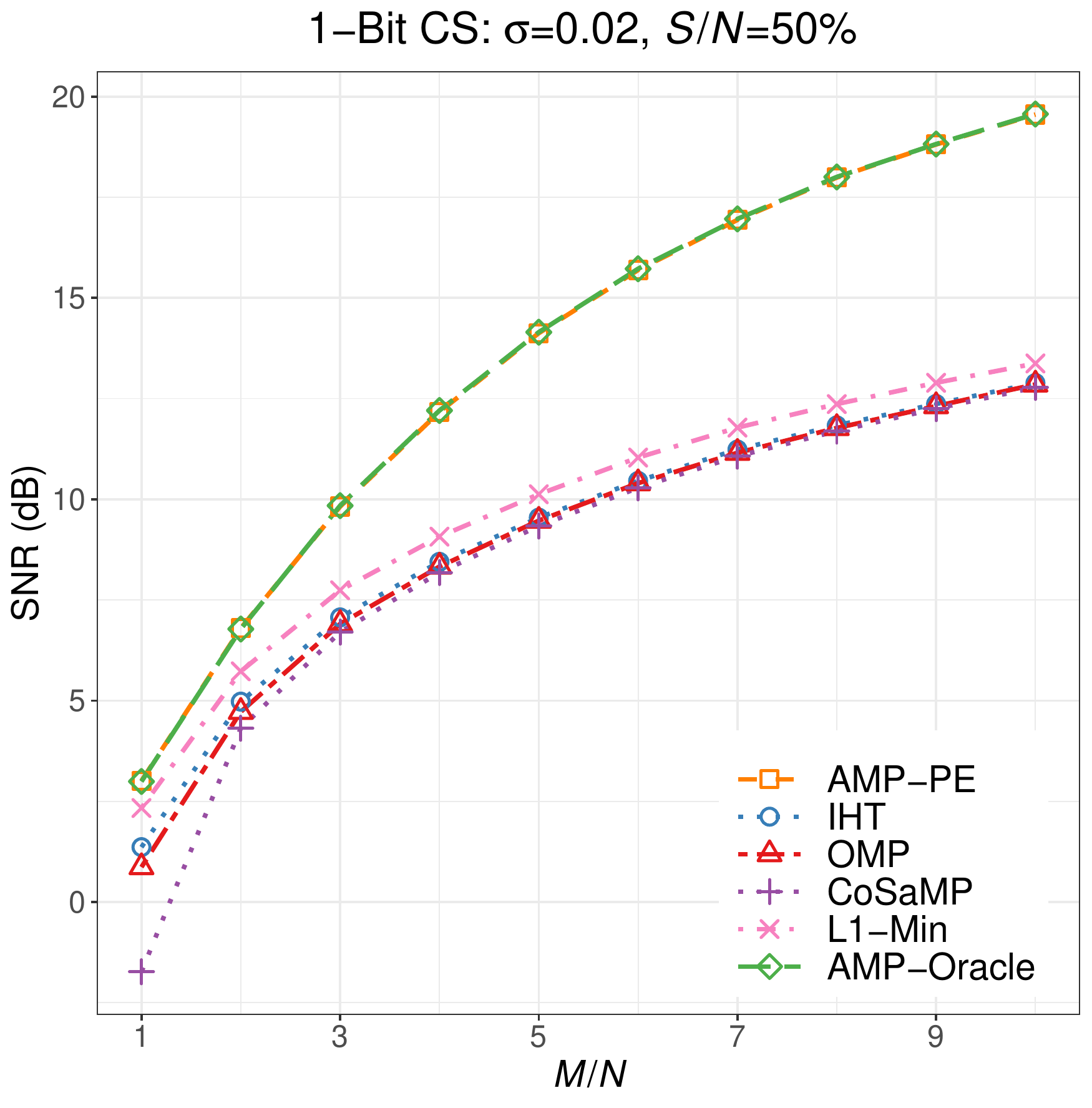}}
\subfigure{
\includegraphics[width=0.3\textwidth]{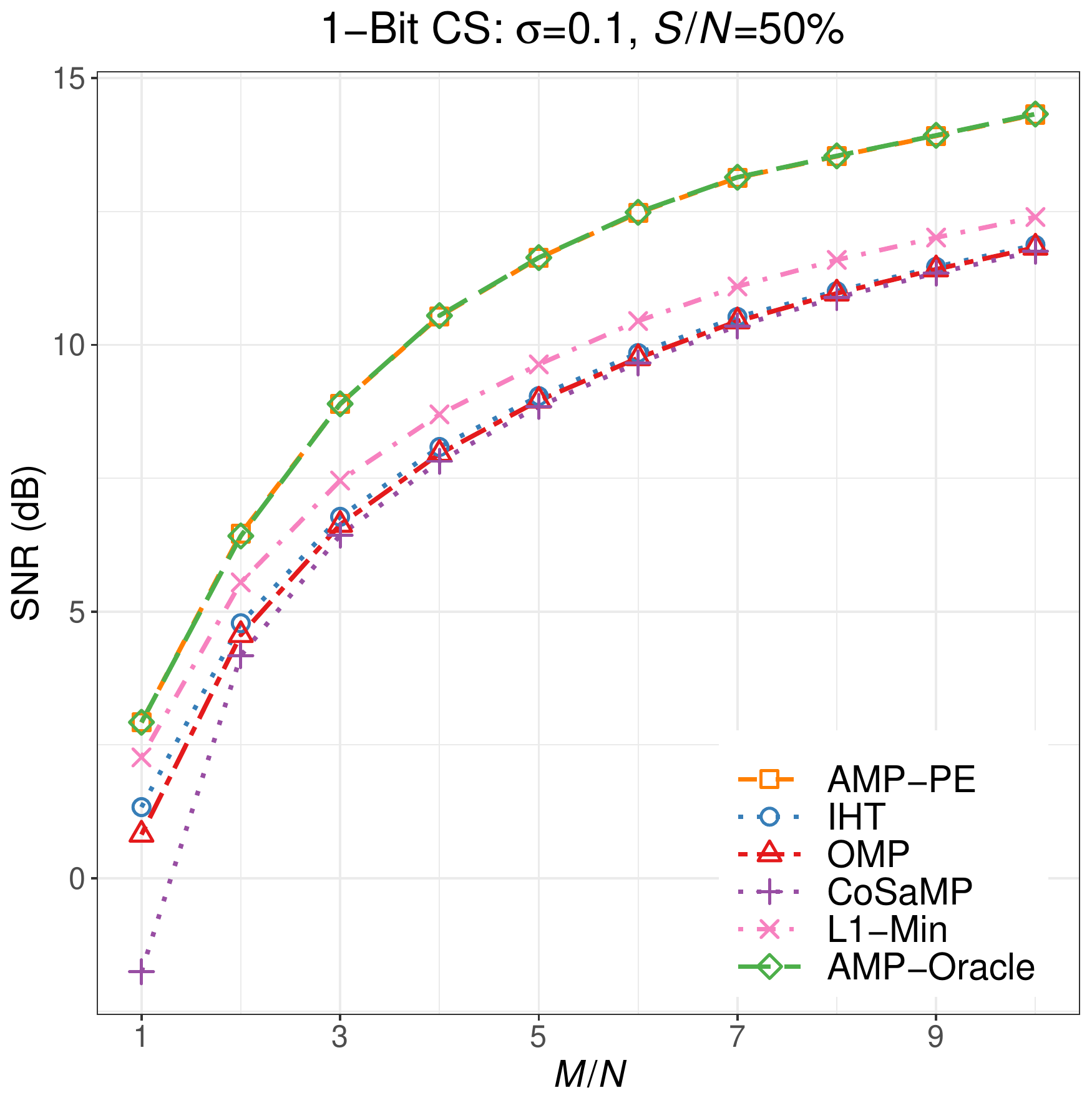}}\\
\subfigure{
\includegraphics[width=0.3\textwidth]{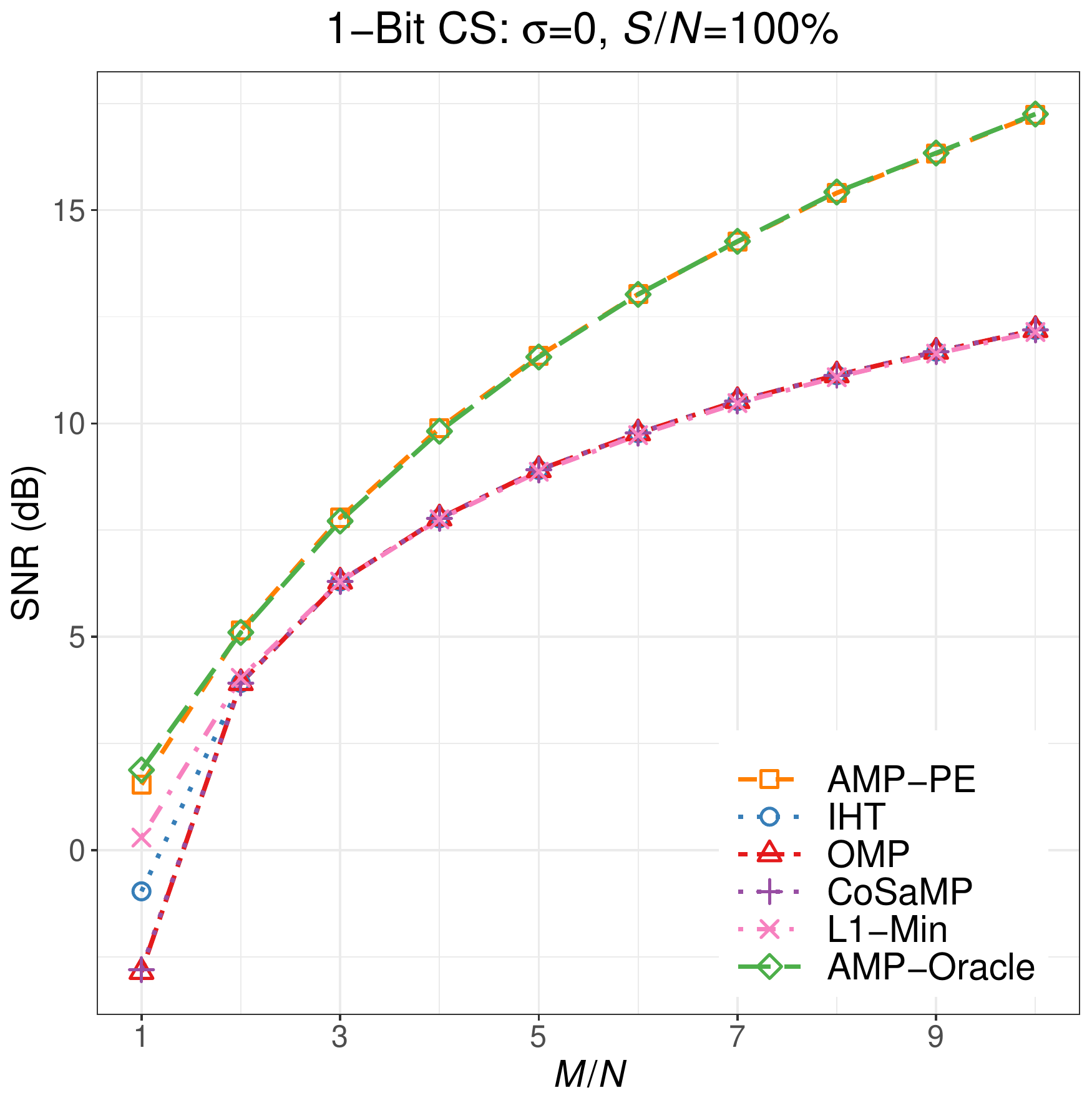}}
\subfigure{
\includegraphics[width=0.3\textwidth]{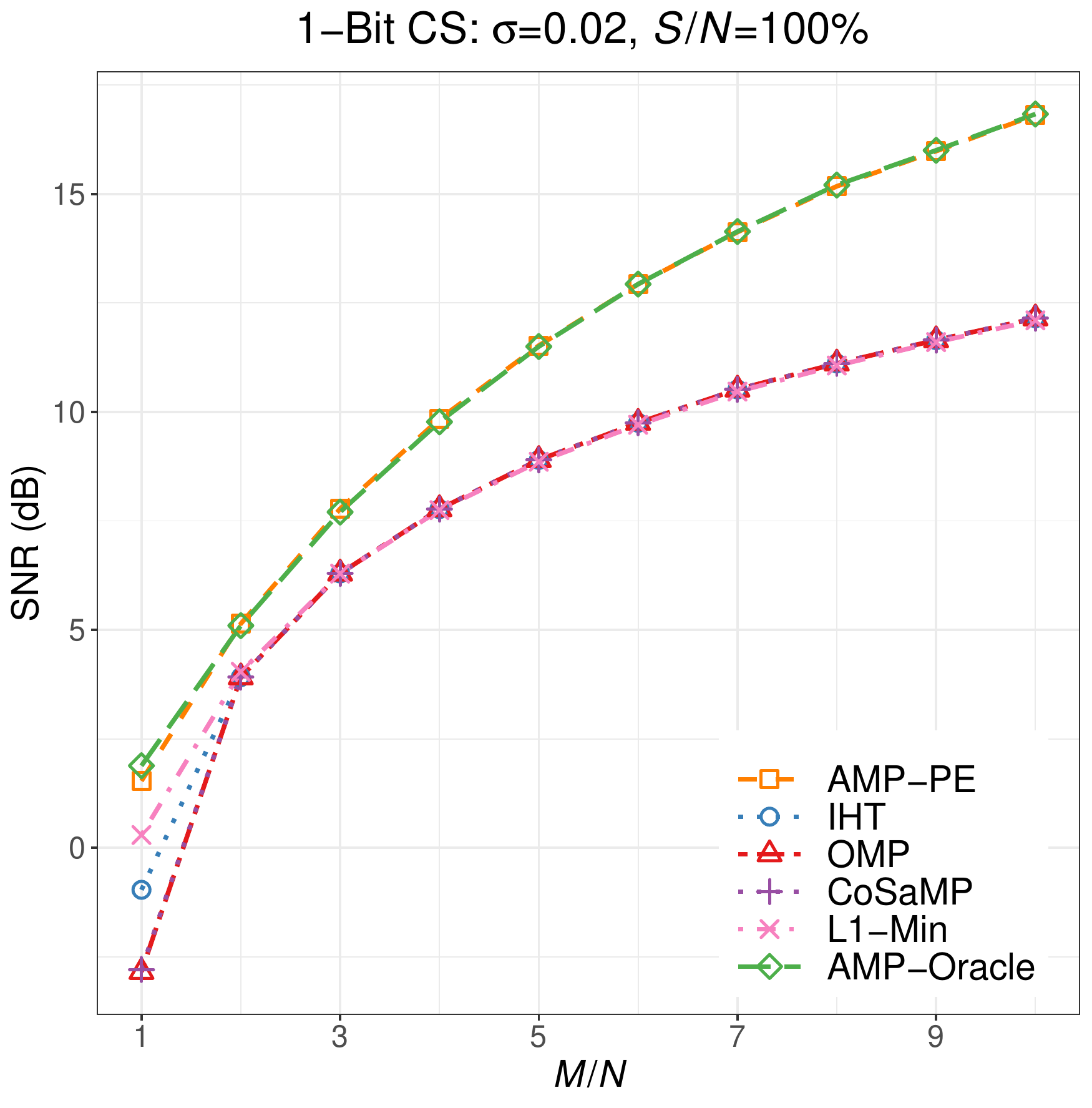}}
\subfigure{
\includegraphics[width=0.3\textwidth]{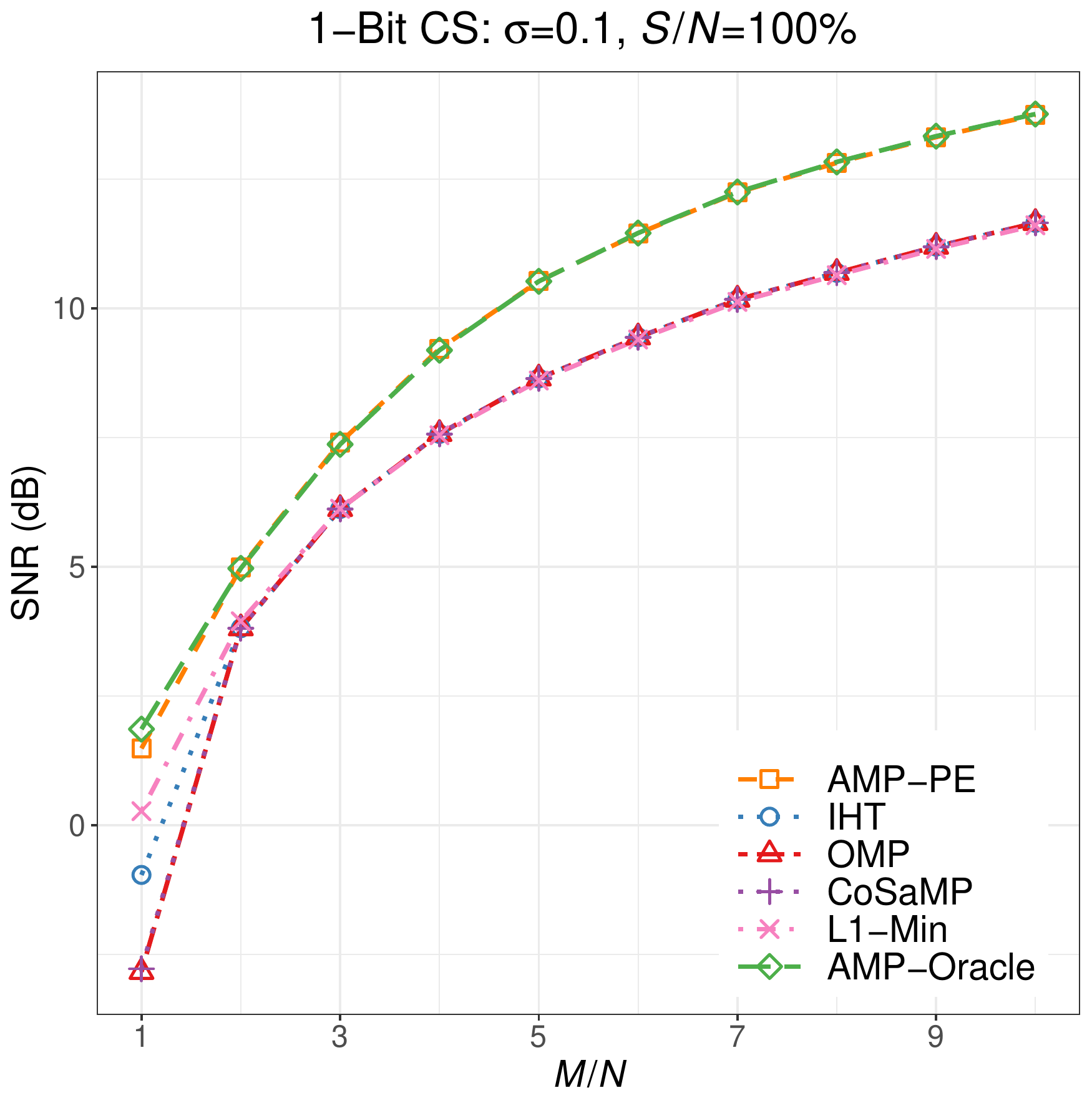}}
\caption{Comparison of different approaches in recovering sparse signals from 1-bit measurements. The sparsity level of the signal $\frac{S}{N}\in\{10\%,50\%,100\%\}$ and the oversampling ratio $\frac{M}{N}\in\{1,2,\cdots,10\}$. The noise $w\sim\mathcal{N}(0,\gamma_w)$ is added before the quantization, with $\gamma_w\in\{0,0.02,0.1\}$ producing zero, moderate, and high levels of pre-quantization noise respectively.}
\label{fig:1bit_experiments}
\end{figure*}

In this section we compare the proposed AMP with built-in parameter estimation (AMP-PE) with the other state-of-the-art sparse recovery methods such as IHT \cite{Blumensath2008,Blumensath2009}, OMP \cite{OMP07,Cai:OMP:2011}, CoSaMP \cite{CoSaMP09}, $l_1$-norm minimization \cite{WaveletAnalysis94,ScaleBasis99}. The proposed AMP-PE jointly recovers the distribution parameters and the signal, whereas the other methods require extensive parameter tuning processes. In the experiments, signals of varying sparsity levels are recovered from quantized measurements under different noise levels. The signal-to-noise ratios (SNR) of the recovered signals are computed to evaluate different methods. Experimental results show that the proposed AMP-PE perform much better than the other methods in the zero and moderate noise regimes, and outperforms the other methods in most of the cases in the high noise regime. Reproducible code and data are available at \urlstyle{tt}\url{https://github.com/shuai-huang/1bit-CS}

For the sparse signal recovery experiments, we fix the signal length $N=1000$ and vary the sparsity level $\frac{S}{N}\in\{10\%,50\%,100\%\}$ and the oversampling ratio $\frac{M}{N}\in\{1,2,\cdots,10\}$, where $S$ is the number of nonzero entries in $\vx$ and $M$ is the number of measurements. Specifically, the nonzero entries are randomly generated from the distribution $\mathcal{N}(0,1)$, and so is the random Gaussian measurement matrix $\mA$. The columns of $\mA$ are further normalized. We add the white Gaussian noise $\vw\sim\mathcal{N}(0,\gamma_w)$ to the noiseless measurements $\vz=\mA\vx$ before quantization, with $\gamma_w\in\{0, 0.02, 0.1\}$ producing zero, moderate, and high levels of pre-quantization noise respectively. The 1-bit measurement $y_m$ is then obtained by applying the quantizer $\mathcal{Q}$: $y_m=\mathcal{Q}(z_m+w_m)$.

For each combination of $\{\frac{S}{N},\frac{M}{N},\gamma_w\}$, the average SNR across 100 random trials is computed for each method, and the results are shown in Fig. \ref{fig:1bit_experiments}. Here we also include the results from AMP using true distribution parameters (AMP-Oracle), which corresponds to the best results that could possibly be obtained using AMP. We can see that the proposed AMP-PE is able to quickly match the performance of the oracle AMP when the oversampling ratio $\frac{M}{N}\geq 2$. All of the methods are initialized with zero solutions. The damping operation with a rate of $0.1$ is applied to the AMP algorithm. For IHT, OMP and CoSaMP, we assume the sparsity level $\frac{S}{N}$ is already known, which would inevitably give them an unfair advantage. For the $l_1$-norm minimization approach, we tune its regularization parameter on a separate training dataset. The proposed AMP-PE does not require parameter tuning, it treats the distribution parameters as unknown variables and jointly recovers them with the signal. Note that the results from GAMP-Oracle are for reference only. We can see that the proposed AMP-PE generally outperforms the other sparse recovery methods by a lot in the zero-noise and moderate-noise regimes where $\gamma_w\in\{0,0.02\}$. As we move to the high-noise regime where $\gamma_w=0.1$, the proposed AMP-PE still outperforms the other methods in most of the cases.

\section{Conclusion}
\label{sec:con}
Taking a probabilistic perspective, we solve the 1-bit CS problem via the proposed AMP framework where the signal and noise distribution parameters are treated as variables and jointly recovered with the signal. This leads to a much simpler way to estimate the parameters by maximizing their posteriors. It allows us to venture into the complicated quantization noise model in 1-bit CS, whereas previous AMP approaches either prespecify (tune) the noise parameter or use an approximated noise model due to the overwhelming complexity. 

A computationally efficient approach that combines EM and the second-order method is introduced to compute the maximizing parameters. Experimental results show that the proposed approach performs much better than the other state-of-the-art sparse recovery methods in the zero and moderate noise regimes. In the high noise regime the performances of different approaches become similar, the proposed approach still outperforms the other methods in most of the cases. 

The stability (or reliability) of an algorithm is a key factor in deciding its adoption in real applications. The AMP algorithm has long been criticized due to its lack of convergence guarantees for general measurement matrices. The damping and mean removal operations have been proposed to alleviate this issue, and are often quite effective. Initialization also plays an important role in both ensuring the convergence and recovering an accurate solution. In our experiments we observed that it is often the parameter initializations rather than the variable initializations that contribute to the algorithm's performance. Previous works usually overlook the influence of the initialization and focus on establishing convergence conditions for a specific class of random matrices. We believe it would be worth pursuing how the initialization affects the algorithm's convergence behavior. Notwithstanding these drawbacks, our method offers an efficient joint recovery of the signal and parameters from a probabilistic perspective, and pushes forward the state-of-the-art performance.

\begin{appendices}
\counterwithin{assumption}{section}
\counterwithin{theorem}{section}
\renewcommand\thetable{\thesection\arabic{table}}
\renewcommand\thefigure{\thesection\arabic{figure}}

\end{appendices}


\ifCLASSOPTIONcaptionsoff
  \newpage
\fi



%
\bibliographystyle{IEEEbib}
\bibliography{refs}

\end{document}